\documentclass{iopart}
\usepackage{graphicx}

\begin{document}

\title[Equivalence between isospectrality and iso-length spectrality]
{Equivalence between isospectrality and iso-length spectrality
for a certain class of planar billiard domains}

\author{Yuichiro Okada \dag and Akira Shudo \ddag}

\address{Department of Physics, Tokyo Metropolitan University, 
Minami-Ohsawa, Hachioji, Tokyo 192-0397, Japan}


\begin{abstract}

Isospectrality of the planar domains 
which are obtained by successive unfolding of a fundamental building block 
is studied in relation to 
iso-length spectrality of the corresponding domains. 
Although an explicit and exact trace formula such 
as Poisson's summation formula or Selberg's trace formula is not known 
to exist for such planar domains, equivalence between isospectrality and 
iso-length spectrality in a certain setting can be proved 
by employing the matrix representation 
of $\lq\lq$transplantation of eigenfunctions". 
As an application of the equivalence, transplantable pairs of domains, 
which are all isospectral pair of planar domains and therefore counter 
examples of Kac's question $\lq\lq$can one hear the shape of a drum?",  
are numerically enumerated and it is found at least up to the domain 
composed of 13 building blocks transplantable pairs coincide with 
those constructed by the method due to Sunada. 

\end{abstract}


\section{Introduction}

A famous question of M. Kac, $\lq\lq$can one hear the shape of a drum?", 
is concerned with isospectrality of the planar domains \cite{KAC}.
The bounded domains $D_1$ and $D_2$ are called isospectral 
if two domains have the same eigenvalue spectrum up to the degree of 
multiplicities, {\it i.e.}, $spec(D_1) = spec(D_2)$, 
where 
\begin{equation}\label{spectrum}
{\rm spec}(D) = \{ \mu_1 \le \mu_2 \le \cdots \mid
\Delta f_i = \mu_if_i \ {\rm in} \ D, \ f_i = 0 \ {\rm on}\ 
\partial D \ \ (i=1,2,\cdots) \}. 
\end{equation}
Kac's question is alternatively stated as $\lq\lq$are the planar 
domains $D_1$ and $D_2$ congruent if $spec(D_1) = spec(D_2)$?". 

Iso-length spectrality of the domains is analogously introduced 
for the length spectrum of closed billiard trajectories on 
the corresponding domains 
$D_1$ and $D_2$. Also for iso-length spectrality
one can pose the same question as Kac's. 
The problem we discuss here is equivalence between 
isospectrality and iso-length spectrality on the planar domains of 
a certain class.  

If either the eigenvalue spectrum or the length spectrum determines 
the shape of the domain uniquely, the equivalence problem becomes obvious 
since a congruent pair of domains trivially provides both isospectrum and 
iso-length spectrum. 
However, it is known that there exists a non-trivial case where 
equivalence between isospectrality and iso-length spectrality is
concluded.
Such a situation was reported by Milnor for the first time \cite{MIL}.
It is known that there exist the self-dual lattices $L_1$ and $L_2$ 
in ${\bf R}^{16}$ which are not congruent 
in the sense that no rotation of ${\bf R}^{16}$ carries $L_1$ to $L_2$, 
nevertheless they own the same length spectrum.
That is, the quotients of ${\bf R}^{16}$ by these lattices,
${\bf R}^{16}/L_1$ and ${\bf R}^{16}/L_2$ are non-isometric 
but iso-length spectral. 
Furthermore its isospectrality can also be derived 
from the exact trace formula which represents duality relation between 
the spectrum of Laplacian 
acting on the flat torus and the corresponding length spectrum 
of closed geodesics.
The flat torus which is defined as 
a quotient ${\bf R}^n / \Gamma$ of ${\bf R}^n$ by a lattice $\Gamma$ of 
rank $n$, the set of eigenvalues of Helmholtz equation $\Delta f = \mu f$ 
is explicitly given as, 
\begin{equation}\label{espec}
\{ 4\pi^2\|\sigma\|^2 \ \mid \ \sigma \in \Gamma^\ast\}.
\end{equation}
Here $\Gamma^\ast$ denotes the dual lattice of $\Gamma$, i.e., 
$\Gamma^\ast = \{ \sigma \in {\bf R}^n \mid \sigma\cdot\gamma\in 
{\bf Z}\quad\forall \gamma \in \Gamma \}$. 
The corresponding length spectrum which is the set of 
closed geodesics on the same torus is expressed just by 
the distance between the origin and each lattice point on $\Gamma$:
\begin{equation}\label{lspec}
\{\|\gamma\|\mid \gamma \in \Gamma\}.
\end{equation}
The two spectra are connected via Poisson's summation formula or Jacobi 
identity:
\begin{equation}
\sum_{\sigma\in\Gamma^\ast}\exp (-4\pi^2\|\sigma\|^2 t) = 
(4\pi t)^{-n/2} vol({\bf R}^n / \Gamma )\sum_{\gamma\in\Gamma}
\exp (-\|\gamma\|^2 / 4t).
\end{equation}
This trace formula immediately leads us to equivalence between 
the isospectrality and iso-length spectrality, that is,  
two flat tori are isospectral if and only if they have 
the same length spectrum.
Milnor's work is not a direct answer to 
the original version of Kac's question because it is not concerned with 
the planar domains. 
But from Milnor's example one learned that there really exist 
non-congruent domains with a common length and eigenvalue spectrum.

In more general cases where two Riemannian 
manifolds have a common finite Riemannian covering, a sufficient condition 
for isospectrality was given by Sunada \cite{SUN}. Let $G$ be a finite group 
which acts freely on a certain compact Riemannian manifold $M$ by isometries. 
Sunada's theorem tells us that the {\it quotients $M/H_1$ and $M/H_2$ are 
isospectral if there exist two subgroups $H_1$ and $H_2$ of $G$, which 
satisfy so-called almost conjugate condition:
\begin{equation}\label{ACC}
^\sharp ([g]\cap H_1) = ^\sharp ([g]\cap H_2)
\end{equation}
for any conjugacy class $[g]$ in G.}  
The proof is based on a sort of trace formula which can be regarded 
as a prototype of the Selberg trace formula \cite{SUN,BRO1}. 
If subgroups $H_1$ and $H_2$ are conjugate in the usual sense, 
then these quotients $M/H_1$ and $M/H_2$ are merely isometric. 
However it had been known that there exist triplets $(G,H_1,H_2)$ 
so that $H_1$ and $H_2$ are almost conjugate but not conjugate in $G$. 
Such triplets were used to construct some isospectral but non-isometric 
pairs of Riemannian surfaces by Buser \cite{BUS} and Brooks and Tse \cite{BT}. 
Sunada also proved that $M/H_1$ and $M/H_2$ 
have the same length spectrum of closed geodesics \cite{SUN}. 
Therefore, in this case, 
iso-length spectrality is derived together with isospectrality 
from the same source. 

In case of planar domains, after a quarter of century, 
Gordon, Webb, and Wolpert \cite{GWW} 
solved Kac's original question also negatively by constructing an isospectral 
but non-congruent pair of planar domains (see Fig. 1). 
For the construction, they used a version of Sunada's theorem, which was  
extended to the orbiford (billiard) setting by B\'erard \cite{BER}.
Since the source of isospectrality was Sunada's theorem, iso-length 
spectrality naturally follows as well. 

However, if there exists an isospectral pair of planar domains which 
are not given by Sunada's construction, 
equivalence of isospectrality and iso-length spectrality cannot 
straightforwardly be discussed.  In fact, no one can exclude such 
a situation because the Sunada's condition gives merely a sufficient condition 
for the isospectrality and also iso-length spectrality.
Nevertheless, as we will show below, it is possible to 
relate them through the notion of so-called {\it transplantation 
of eigenfunctions}.  The transplantation method is originally invented 
by Buser, Conway, Doyle, and Semmler to 
reexamine isospectrality of some given domains \cite{BCDS}. 
They constructed 17 families of isospectral pairs of domains 
based on the Sunada's method, and confirmed these isospectrality 
by the transplantation method. 
As far as planar domains are concerned, 
these pairs are, to authors' knowledge, all isospectral pairs ever known. 

It should be remarked that the transplantation method is not independent of 
the Sunada's theorem. 
In the above case of quotient manifolds, 
Brooks gave an alternative proof of Sunada's theorem based on the transplantation method, 
which appears a necessary condition for the assumption of the theorem \cite{BRO2} . 
However equivalence between these two methods has not been proved 
in case of planar domains.

\begin{figure}[h]
\begin{center}
\includegraphics[width=.80\linewidth]{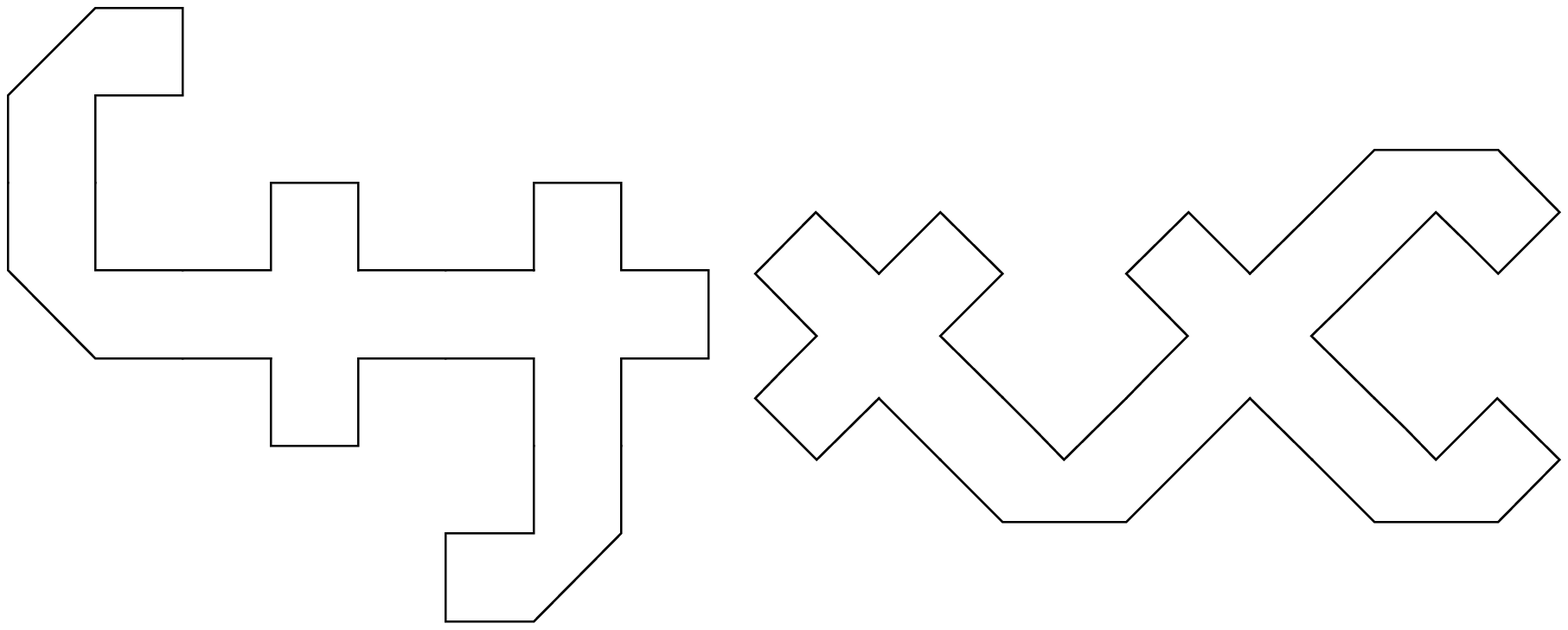}
\end{center}
\caption{Isospectral pairs of domains constructed by Gordon \etal.}
\label{Fig:FIG1}
\end{figure}

The organization of this paper is as follows.
We first explain the transplantation method in section 2 and 
give its matrix representation in section 3. 
In section 4, we provide a proof showing that 
isospectrality and iso-length spectrality is equivalent for a certain   
class of domains, that is, those domains constructed by 
successive unfolding of a fundamental building block. 
All isospectral pairs ever known belong to this class. 
In section 5 we enumerate isospectral pairs of domains and 
compared them with the table obtained by 
Buser, Conway, Doyle, and Semmler \cite{BCDS}. 
It is shown that our labor of searching isospectral pairs is considerably 
reduced as a bi-product of our result in section 4.  Section 6 is 
devoted to summary and discussion of our work.


\section{Transplantation method}

Let us consider two domains $D_1$ and $D_2$ as Fig. 2. 
We here describe the transplantation method by taking an example of 
the proof of isospectrality based on it.
We follow the exposition given by Buser \etal \cite{BCDS}. 

Transplantation is a procedure which is
cutting an arbitrary function $f$ 
defined on $D_1$ into its restrictions on each building block, 
say $f_1,f_2,\ldots,f_7$, 
and rebuilding a new function $g$ on $D_2$ by superposing $f_1,f_2,\ldots,f_7$.
Note that such a procedure relied crucially on the peculiarity of the domains,
which are composed of several pieces obtained by 
successive reflection of a certain common building block.  


By the transplantation shown in Fig. 2, any eigenfunction $f$ 
with any eigenvalue $\mu$ on $D_1$ is transplanted onto $D_2$ 
so that the transform $g$ is also an eigenfunction with the same eigenvalue 
$\mu$. This can actually be confirmed by checking 
the following conditions:

\begin{itemize}
\item {\it $\Delta g=\mu g$ in the interior of each piece.}\\ 
This is always true because $g$ is given as a sum of several $f_i$'s 
on each piece.
\item {\it The function g is smoothly connected on all reflecting segments}\\ 
This is not trivial, but true on this example. On the reflecting segment $r$ 
in Fig. 2, for example, $-f_2+f_3-f_5$ fits smoothly with $+f_2+f_6-f_7$ 
because $f_3$ and $f_6$, or $-f_5$ and $-f_7$ was smoothly connected 
on the original domain $D_1$ 
and $-f_2$ and $f_2$ also fits smoothly on the segment 
by the reflection principle.  We can also check the smoothness condition 
on the other reflecting segments in the same way.
\item {\it The function g vanishes on the boundary of }$D_2$. \\ 
This is not trivial, but true on this example. On the boundary segment $b$ 
in Fig. 2, for example, $-f_2+f_3-f_5$ vanishes  because $-f_2$ 
cancels $f_3$ there and $f_5$ was already zero.  
We can also check the boundary condition on the other boundary segments 
in the same way.
\end{itemize}
One more condition is invertibility of transplantation, which 
guarantees that the dimensions of the 
space belonging to the eigenvalue $\mu$, {\it i.e.},  
the multiplicity of $\mu$, 
on $D_1$ is equal to the dimension on $D_2$ or less. 
This is expressed as the condition,
\begin{itemize}
\item {\it The transplantation is invertible as a linear map.}\\ 
This is not trivial, 
but can be easily checked to be true for this example.
\end{itemize}
Thus these four conditions lead to $Spec(D_1)\subset Spec(D_2)$.
The last condition becomes the check to see that any eigenfunction 
with any eigenvalue $\nu$ on $D_2$ can be transplanted onto $D_1$
conversely, that is, $spec(D_2)\subset spec(D_1)$.
In this way, we know that these two domains are isospectral.

\begin{figure}[h]
\begin{center}
\includegraphics[width=.80\linewidth]{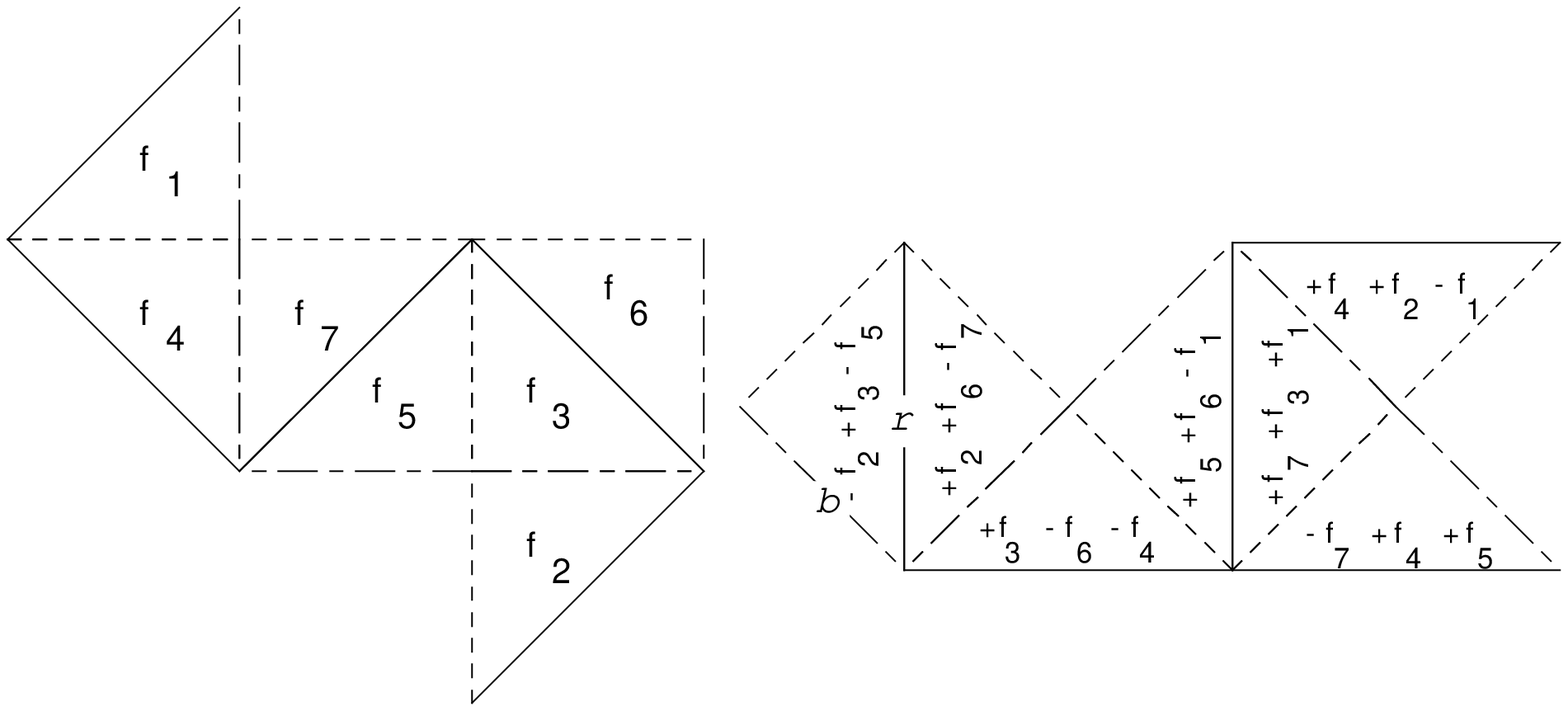}
\end{center}
\caption{Isospectral pairs of domains and the transplantation. }
\label{Fig:GWW}
\end{figure}

These are conditions for eigenfunctions satisfying Dirichlet 
boundary conditions, but they are also Neumann isospectral. 
This can be verified 
by changing all the minus signs before each element of
transplanted function $f_i$ to the plus one 
when one superposes them on the transformed domain.
 Conversely it is possible to construct Dirichlet isospectral pairs 
once Neumann isospectral are at hand. 
The procedure is as follows: (i) for both domains, 
attach the code $\sigma_i =$ 1 or -1 to each building block $i$ 
such that the codes of neighboring blocks are different, 
(ii) let the building block $i$ with code $\sigma_i$ be 
transformed onto the building block $i'$ with code $\sigma_i'$, 
and if $\sigma_i \sigma_{i'}$ = 1 then the sign before $f_i$ is set to 
be plus and otherwise minus.
It is easy to show that the resulting pair of domains are Dirichlet isospectral. 
Therefore,  
isospectrality with respect to the  Dirichlet and Neumann 
boundary conditions is equivalent. 

We also remark that the above proof is independent 
of the choice of a building block. 
This means that this example represents not only a single pair of 
isospectral domains but also a family of isospectral pairs, which 
are realized by replacing a building triangle simultaneously 
with another shaped building block. 
In particular, we can obtain the example of Gordon \etal. 
in Fig. 1 by replacing the building block in Fig. 2 by a heptagon.


\section{Transplantability and its matrix representation}

In this section, we provide an alternative and more transparent 
condition for the {\it transplantability} of eigenfunctions. 
In what follows, we limit ourselves to the class of domains 
constructed by successive reflection of a fundamental building block,
which we call {\it unfolded domains}, since transplantation of 
eigenfunctions is most simply done between such domains. 
In order to make successive unfolding 
possible, each building block is supposed to have three line segments 
at which the building block is reflected.
The domains shown in Figs 1 and 2 are the examples.
Hereafter we denote $N$ as the number of building blocks 
composing the unfolded domain.  

The whole shape of the unfolded domain 
is determined by the choice of the building block and the rule 
specifying unfolding. The unfolding rule can be identified with 
the edge-colored graph and the correspondence is given in table 1. 
Each piece obtained by reflection is represented by the 
vertex of the graph, and the reflecting segment by its edge. 
The three boundary segments of the building block
are distinguished by the color of the edges. 
Some boundary segments of the building block form the boundary of 
the whole unfolded domain, and we represent 
the edge corresponding to one of such boundary segments as the closed loop. 
An example of edge-colored graph is demonstrated in Fig. 3.

\begin{table}[h]
\caption{}
\begin{indented}
\item[]
\begin{tabular}{ccc}
\br
unfolding rule &  & edge-colored graph\\
\mr
pieces & $\longleftrightarrow$ & vertices\\
reflecting segments & $\longleftrightarrow$ & edges\\
kind of segment & $\longleftrightarrow$ & color of edge\\
\br
\end{tabular}
\end{indented}
\end{table}

\begin{figure}[h]
\begin{center}
\includegraphics[width=.80\linewidth]{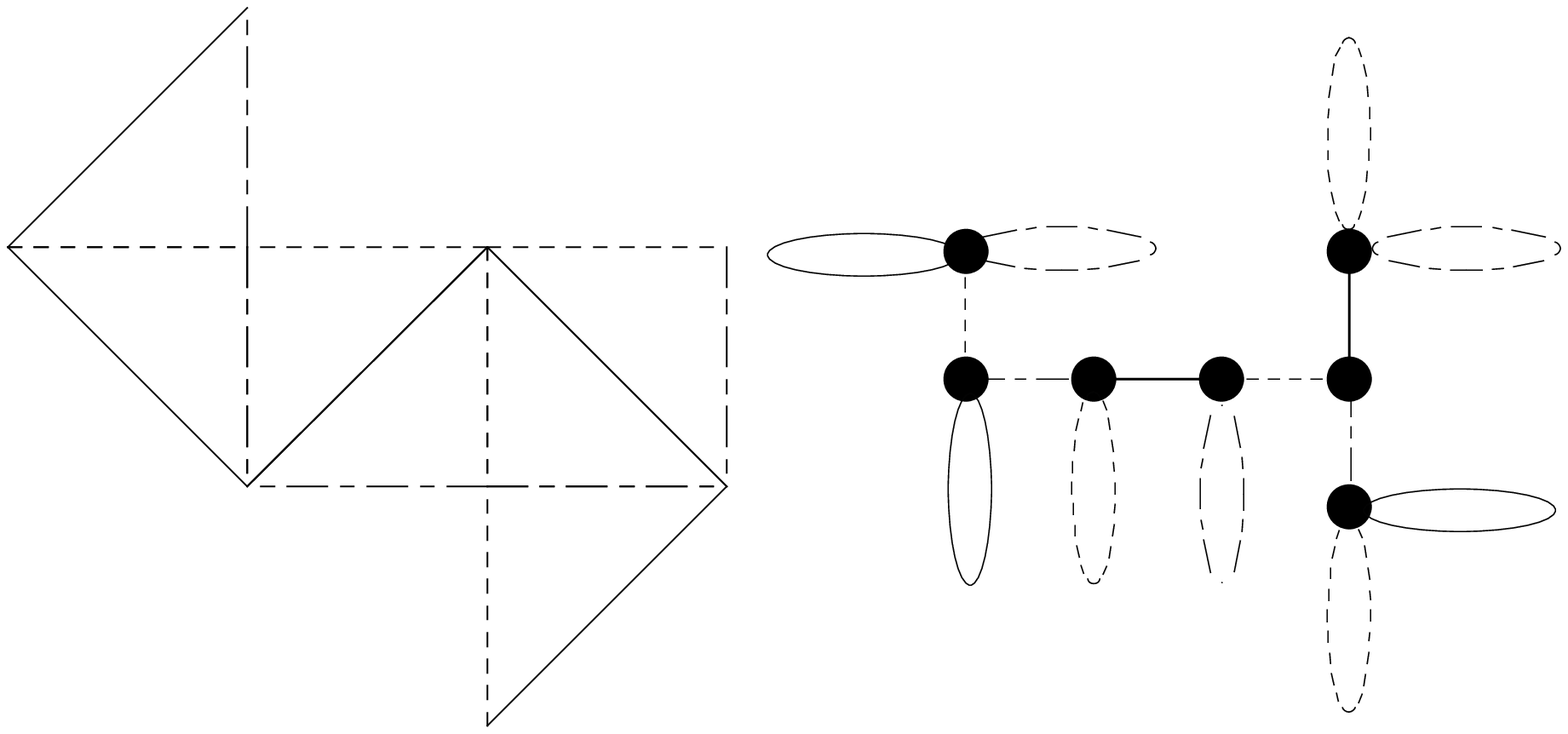}
\end{center}
\caption{An example of edge-colored graph representing an unfolding rule.
The three types of lines(solid, dotted, broken with dots) denote 
three colors.}
\label{Fig:FIG3}
\end{figure}

The structure of a graph is also represented by the adjacency matrix. 
That is, the reflecting rule to generate an unfolded domain $P$ is given  
by the triplet of adjacency matrices $M^P_a$, $M^P_b$, and $M^P_c$ defined as
follows:
\begin{equation}
{(M^P_\gamma)}_{ij}=
\cases{1 \ \ & \mbox{if the piece $i$ is adjacent to $j$ by segment $\gamma$}\\
0 \ \ & \mbox{otherwise}}.
\end{equation}
$$
\gamma = a,b,c.
$$
Diagonal element ${(M^P_\gamma)}_{ii}$ is set to be 1 if the piece $i$ 
has the boundary segment $\gamma$. The adjacency matrix is, in general, 
not uniquely determined even if one fixed the graph 
because there remains a freedom to determine the order of labeling vertices. 
In this setting, two unfolded domains $P$ 
and $Q$ are congruent if there exists a permutation matrix $U$ such that 
\begin{equation}\label{congruent}
U M^P_a U^{-1}=M^Q_a,\quad U M^P_b U^{-1}=M^Q_b,\quad U M^P_c U^{-1}=M^Q_c.
\end{equation}
Here the permutation matrix is defined as a $N \times N$ matrix in which 
each row or column has only one unity, and all the other elements are zero.
If the building block has some symmetry, for example reflecting symmetry 
with respect to a certain axis, 
it might happen that non-permutation matrix connects the congruent pair 
$P$ and $Q$. However, if one destructs the symmetry by changing 
the shape of the building block, such an $\lq\lq$accidental" situation can be 
avoided. The condition (\ref{congruent}) for a permutation 
matrix $U$, therefore, becomes the necessary and 
sufficient condition for the congruency of $P$ and $Q$ in this sense. 

Next we give the condition for transplantability of 
eigenfunction of two given unfolded domains using the adjacency matrices.
By transplantability we mean the possibility that isospectrality 
of a pair of unfolded domains can be verified by the transplantation method. 

The transplantation procedure described in section 3 is explicitly 
expressed as follows. Let $P$ and $Q$ be two unfolded domains 
composed of $N$ copies of a common building block. 
Let $T:L^2(P)\ni f \mapsto g \in L^2(Q)$ be a linear transformation. 
We say that the transformation $T$ is a {\it transplantation} 
if the transform $g$ is obtained by cutting and pasting $f$, namely,
\begin{equation}
g_i = \sum_{j=1}^{N}T_{ij} f_j,\quad i=1,2,\ldots ,N, 
\end{equation}
where $f_j$ and $g_i$ denote restrictions of $f$ and $g=Tf$ to each piece, 
respectively. Also as mentioned in section 2, 
it  is not trivial at all that the transform $g$ becomes an eigenfunction 
on the transformed domain at this stage. 

For given unfolded domains $P$ and $Q$, 
we say that they are {\it transplantable} 
if there exists an invertible transplantation $T$ satisfying  
the following conditions: 
\begin{equation}\label{transplantable}
\quad T M^P_a T^{-1}= M^Q_a,\quad T M^P_b T^{-1} 
= M^Q_b,\quad T M^P_c T^{-1}= M^Q_c.
\end{equation}
It can be easily checked that 
smoothness on all the reflecting segments and the 
boundary conditions on the unfolded domains 
are satisfied if the above conditions are fulfilled.
Here, existence of the inverse matrix $T^{-1}$ corresponds to 
the invertible condition. The condition (\ref{transplantable}) 
is the one for transplantability of Neumann boundary condition, but 
as remarked in section 2, it is equivalent to 
transplantability of the Dirichlet case.
As was seen in the previous section, $P$ and $Q$ are isospectral 
if they are transplantable. 

Note that if $T$ is a permutation matrix, 
then $P$ and $Q$ are merely congruent. Therefore, in order to 
construct isospectral but non-congruent pair of unfolded domains, 
we must find a matrix satisfying the condition (\ref{transplantable}), 
but not a permutation matrix.


\section{Transplantability and iso-length spectrality}

We next consider the length spectrum of the unfolded domain.
The length spectrum is the set of length of closed trajectories or 
periodic orbits on the 
billiard domain. We first consider the relation between 
closed trajectories on the unfolded domain $P$ and its fundamental 
building block $B$.
Note that any periodic orbit on $P$ can be regarded as 
the lift of a closed trajectory on $B$, because its projection is always
a periodic orbit on $B$. However, the converse is not necessarily true.
A single closed orbit on $B$ yields $N$  trajectories as 
its lift on the unfolded domain $P$, which are sometimes closed 
orbits on $P$ but the others not.  
By the construction of the adjacency matrix  
$M^P_{\gamma}$, it is easy to see that the number of closed lifts 
of a given closed trajectory on $B$ is counted as 
\begin{equation}\label{NUMBER}
n^P(\gamma)=\Tr ( M^P_{\gamma_m} M^P_{\gamma_{m-1}} \cdots M^P_{\gamma_1}),
\end{equation}
where $\gamma=\gamma_1 \gamma_2 \cdots \gamma_m \ (\gamma_i = a, b,c )$ 
denotes the sequence representing in which order a given closed trajectory 
on $B$ hits the boundary segments (for example, see Fig. 4).
Such a sequence is not determined uniquely for a given closed orbit. 
In fact, 
if $\gamma=\gamma_1 \gamma_2 \cdots \gamma_m$ is a sequence 
specifying some closed trajectory, then all sequences obtained by its 
cyclic shift 
such as $\gamma_2 \cdots \gamma_m \gamma_1$, 
$\gamma_3 \cdots \gamma_m \gamma_1 \gamma_2$, {\it etc.} 
can be regarded as the sequence giving to the same closed trajectory. 
However the number of closed lifts is invariant because of the relation:
\begin{equation}
\eqalign{
\Tr ( M^P_{\gamma_m} \cdots M^P_{\gamma_2} M^P_{\gamma_1})
&=\Tr (M^P_{\gamma_1} M^P_{\gamma_m} \cdots M^P_{\gamma_2} M^P_{\gamma_1} 
({M^P_{\gamma_1}})^{-1})\\
&=\Tr ( M^P_{\gamma_1} M^P_{\gamma_m} \cdots M^P_{\gamma_2})\\
&=\cdots .}
\end{equation}
Thus the length spectrum of closed trajectories for  
$P$ is determined by the length spectrum for  $B$ and $n^P(\gamma)$. 

\begin{figure}[h]
\begin{center}
\includegraphics[width=.80\linewidth]{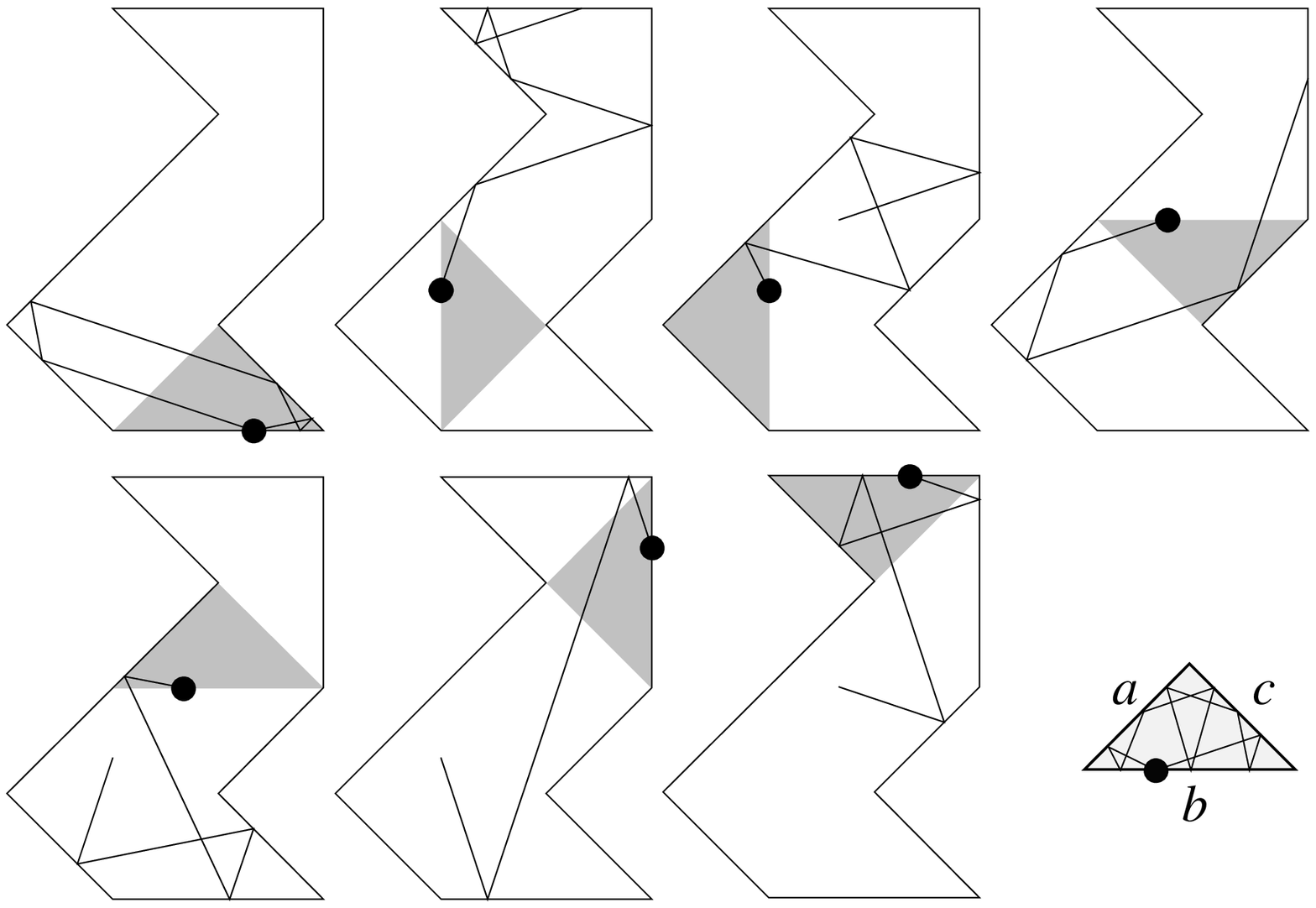}
\end{center}
\caption{Lifts of a closed billiard trajectory $o$ with the sequence 
$\gamma=babacbacbc$. For the sequence $\gamma$, $n^P(\gamma)$ is equal to 1.}
\label{Fig:LIFT}
\end{figure}

In order that two unfolded domains $P$ and $Q$ constructed by the 
same-shaped building block $B$ 
have the same length spectrum, it is sufficient that the numbers of 
closed lifts should coincide for all possible sequences of $\gamma$. 
It can easily be  checked that $n^P(\gamma)$ and $n^Q(\gamma)$ are 
invariant if $P$ and $Q$ are transplantable, {\it i.e., } 
the condition (\ref{transplantable}) is satisfied for the adjacent matrices 
of $P$ and $Q$. Therefore we obtain the following claim.\bigskip

\noindent{\bf Proposition 1}\quad
{\it Let $P$ and $Q$ be two unfolded domains obtained 
by N times successive reflections of the same building block . 
If $P$ and $Q$ are transplantable, 
then $n^P(\gamma)=n^Q(\gamma) \mbox{ for any sequence $\gamma$}$, 
which leads to iso-length spectrality of $P$ and $Q$.}\bigskip

The inverse of the above proposition is to inquire whether or not 
 $n^P(\gamma)=n^Q(\gamma)$ gives the sufficient condition for 
transplantability.  The following statement gives its answer:  
the condition $n^P(\gamma)=n^Q(\gamma)$ is precisely the analog of the 
transplantability of eigenfunctions. \bigskip

\noindent{\bf Proposition 2}\quad
{\it Let $P$ and $Q$ be two unfolded domains obtained 
by N times successive reflection of the same building block. 
If $n^P(\gamma)=n^Q(\gamma)\quad\mbox{for any sequence $\gamma$}$, 
then $P$ and $Q$ are transplantable, which leads to isospectrality 
of $P$ and $Q$.}\bigskip

\noindent{\it Proof}\quad
We have shown that transplantability implies isospectrality in section 3. 
Here we will show that transplantability follows 
from $n^P(\gamma)=n^Q(\gamma)$. 
Let $G^P$ and $G^Q$ be the groups generated by adjacency matrices:
\begin{equation}
G^P = \langle M^P_a, M^P_b, M^P_c \rangle ,\quad
G^Q = \langle M^Q_a, M^Q_b, M^Q_c \rangle.
\end{equation}
Since adjacency matrices themselves are also permutation matrices introduced 
in section 4, $G^P$ and $G^Q$ are subgroups of the symmetric group $S_N$, 
so they are finite. 
Let $F_3$ be a free group generated by characters $a$, $b$, and $c$. 
We can define surjective homomorphism:
\begin{equation}
\Phi^P : F_3 \ni \gamma_1 \gamma_2 \ldots \gamma_m \longmapsto M^P_{\gamma_m} 
M^P_{\gamma_{m-1}} \cdots M^P_{\gamma_1}\in G^P,
\end{equation}
then $G^P$ is isomorphic to $F_3 / \ker \Phi^P$. $G^Q$ is also isomorphic 
to $F_3 / \ker \Phi^Q$.

We now assume that $n^P(\gamma)=n^Q(\gamma)$ for any sequence $\gamma$, 
and note that $M_{\gamma}^P=E \Leftrightarrow n_P(\gamma)=N$, where 
$E$ denotes $N\times N$ identity matrix.
Then we have 
\begin{equation}
\eqalign{
\ker \Phi^P&=\{\gamma \mid M^P_\gamma=E\}\\
&=\{\gamma \mid n_P(\gamma)=N\}\\
&=\{\gamma \mid n_Q(\gamma)=N\}\\
&=\{\gamma \mid M^Q_\gamma=E\}.\\
&=\ker \Phi^Q}
\end{equation}
Here $M^P_\gamma$ and $M^Q_\gamma$ denote $M^P_{\gamma_m} M^P_{\gamma_{m-1}} 
\cdots M^P_{\gamma_1}$ and $M^Q_{\gamma_m} M^Q_{\gamma_{m-1}} 
\cdots M^Q_{\gamma_1}$, respectively. 
This means that $G^P$ is isomorphic to $G^Q$. We can suppose identity maps: 
\begin{equation}
\rho^P: G^P \rightarrow GL(N,{\bf C}),\quad
\rho^Q: G^Q \rightarrow GL(N,{\bf C})
\end{equation}
to be linear representations of $G^P$ and $G^Q$, respectively. 
Since $G^P$ is isomorphic to $G^Q$, 
\begin{equation}
\rho: G^P\ni M^P_\gamma \mapsto M^Q_\gamma \in GL(N,{\bf C})
\end{equation}
is another linear representation of $G^P$. Note that $n^P(\gamma)$ 
and $n^Q(\gamma)$ become the characters of representations $\rho^P$ 
and $\rho$ respectively, and they are equal by our assumption.
 Character theory tells that two representations 
 with the same character are similar\cite{SER}. 
 This means that there exists an invertible matrix $T$ such that
\begin{equation}
T M^P_\gamma = M^Q_\gamma T \quad \mbox{for any $\gamma$}
\end{equation}
This leads to transplantability between $P$ and $Q$. 
\hspace{40mm}(Q.E.D)

By propositions 1 and 2, we know equivalence between transplantability 
and coincidence of the numbers of closed lifts, 
which are sufficient conditions for isospectrality 
and iso-length spectrality respectively. 
However if we assume transplantability, which is more strict 
condition than isospectrality, 
then iso-length spectrality follows. 
Conversely, we assume coincidence of the numbers of closed lifts, 
as a sufficient condition for iso-length spectrality, 
then isospectrality follows. Our results can be summarized 
as the following schema.

\begin{eqnarray*}
\mbox{isospectrality}&\Leftarrow&\mbox{transplantability}\\
&\Leftrightarrow& \forall \gamma \quad n^P(\gamma)=n^Q(\gamma)\\
&\Rightarrow& \mbox{iso-length spectrality}
\end{eqnarray*}


\section{List of transplantable pairs for $N\le 13$}

As an application of the above propositions, we enumerate 
isospectral pairs of unfolded domains whose isospectrality can be verified 
by the transplantation method. 
The candidate space we have to explore is the set ${\cal U}_N$ 
which consists of all possible edge-colored graphs without cycles,
each of which represents an unfolding rule.
The absence of cycles guarantees for the unfolded domain to be formed flatly
for any shaped building block.

The advantage to use the propositions 1 and 2 (and their proof) 
is that even for the purpose of listing up 
transplantable pairs(or triplets if possible), 
we do not have to make pairwise comparisons. 
This is because we can regard each relation $n^P(\gamma)=n^Q(\gamma)$ 
as a necessary condition for the transplantability of $P$ and $Q$,
and use each relation as a $\lq\lq$filter" to pick up transplantable candidates.
More precise steps to find out the transplantable pairs is as follows:
\begin{enumerate}
\item For all edge-colored graphs $P$ in ${\cal U}_N$, compute the character 
$n^P(\gamma)$ for all possible $\gamma$. 
\item prepare the spectrum of characters $\{n^P(\gamma)\}$ by ordering 
the character $n^P(\gamma)$ according to an appropriate rule for $\gamma$
(lexicographical order of sequence $\gamma$, for example). 
\item pick up the pair of edge-colored graphs whose character spectra 
are the same. 
\end{enumerate}
According to the proposition 2, if  one finds the identical 
character spectra, {\it i.e.,} $\{n^P(\gamma)\}$ = $\{ n^Q(\gamma)\}$, 
they yield the transplantable pairs. 
One might be afraid that the step (i), {\it i.e.,} 
enumeration of the character spectrum seems to require infinite 
number of steps, reflecting the fact that  $\gamma$ should run over 
all possible sequences composed of $a,b,c$.  However, this is not the case, 
since 
$G^P$ only forms a finite group as mentioned in the proof of proposition 2. 

It is interesting to note that coincidence of the 
$\lq\lq$ground state", {\it i.e.},
$n^P(\gamma) =n^Q(\gamma)$ for null sequence of $\gamma$ leads that 
the number of building blocks forming the unfolded domains is equal, which 
can be read that the area of two domains are equal. 
Also the condition for the $\lq\lq$first exited state", {\it i.e.},
$n^P(\gamma) =n^Q(\gamma)$ where $\gamma = a, b\ {\rm or}\ c$, 
gives the condition for the length of the boundary to coincide. 
These geometrical quantities agree with the first and second 
terms in the Weyl expansion respectively \cite{KAC,BS,MS}. 

\begin{table}[h]
\caption{The list of transplantable pairs up to $N\le 13$}
\begin{indented}
\item[]
\begin{tabular}{cccccc}
\br
$N$ & $^\sharp{\cal U}_N$ & &
$N$ & $^\sharp{\cal U}_N$ & \\
\mr
2 & 3 & none & 8 & 450 & none\\
3 & 3 & none & 9 & 1326 & none\\
4 & 10 & none & 10& 4262 & none\\
5 & 18 & none & 11& 13566 & none\\
6 & 57 & none & 12& 44772 & none\\
7 & 143& 7 pairs & 13& 148577 & 26 pairs\\
\br
\end{tabular}
\end{indented}
\end{table}

\begin{figure}[h]
\begin{center}
\includegraphics[width=.90\linewidth]{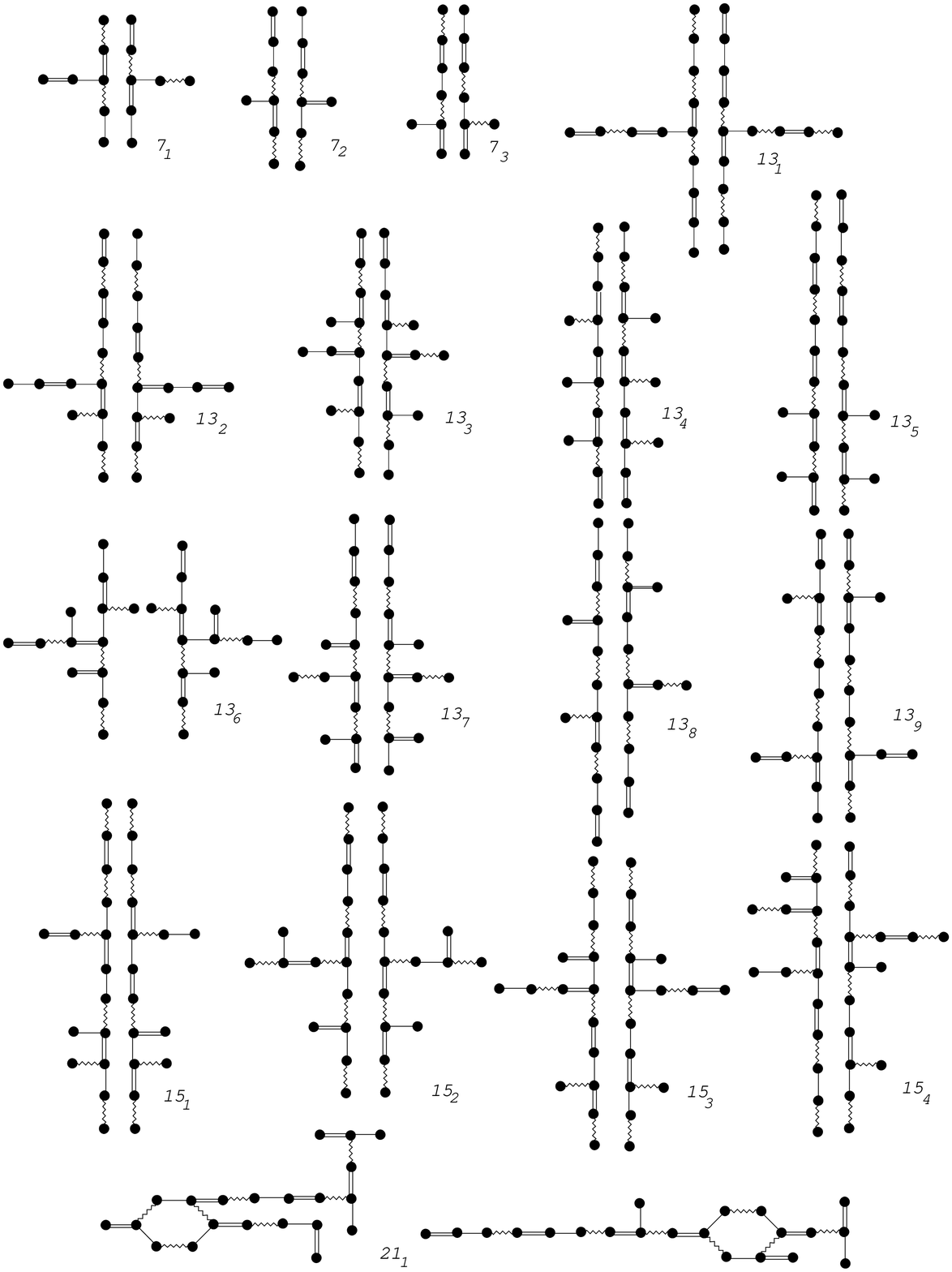}
\end{center}
\caption{Known isospectral pairs of domains
represented as edge-colored graph. 
Three colors of edges are distinguished by the types of lines
(solid, zig-zag, double), and their loops are omitted.
Note that if the colors of edges are permuted the resulting 
pairs also become isospectral.
For example, the pair $7_2$ represents three distinguishable pairs of domains
for a fixed building block.}
\label{Fig:FIG5}
\end{figure}

Applying the above procedure to the edge colored graphs up to $N=13$, 
we have enumerated all possible transplantable pairs, which 
are listed in Table 2.
Fig. 5 shows isospectral pairs ever known, which are listed in  \cite{BCDS}. 
Our result completely 
agrees with the list of known transplantable pairs presented 
in Fig. 5 at least up to $N = 13$.  
Note that all 17 examples given in \cite{BCDS} have been generated 
based on the Sunada's method.  



\section{Summary and discussion}
We have shown that transplantability of eigenfunctions on 
unfolded domains is equivalent to coincidence of 
the character spectrum $\{ n(\gamma)\}$. 
The former is a sufficient condition for 
isospectrality and the latter for iso-length spectrality.

As mentioned in the introduction, if so-called {\it almost conjugate 
condition} (\ref{ACC}) is satisfied for a pair of subgroups $H_1$ and $H_2$, 
one knows that the quotients $M/H_1$ and $M/H_2$ are isospectral, and 
iso-length spectrality of $M/H_1$ and $M/H_2$ simultaneously follows. 
The first counter example of Kac's original problem 
has been constructed with this Sunada's condition \cite{GWW}. 
Once an isospectral pair is obtained based on such a group 
theoretical argument together with the orbifold construction, 
one can easily check isospectrality of the resulting pair 
using the transplantation method. In this way, at the beginning, 
transplantation procedure serves only as a check or 
demonstration for a given pair of domains to be really isospectral.
It is uncertain whether or not there exist transplantable 
pairs of domains which are not constructed based on the Sunada's condition, 
while the Sunada's condition is a sufficient condition for 
transplantability. 

However, by expressing an explicit condition (\ref{ACC}) for transplantability 
using adjacency matrices, 
one notices the similarity between the almost conjugate 
condition (\ref{ACC}) in Sunada's theorem and 
the condition for existence of non permutation matrix $T$ 
in (\ref{transplantable}).  Even if there exists an invertible 
matrix $T$ satisfying (\ref{transplantable}), it is not necessarily 
true that they give an isospectral pair because they are congruent 
if $T$ is merely a permutation matrix. 
Such a situation reminds us of the case where the almost conjugate 
condition (\ref{ACC}) is satisfied but the subgroups $H_1$ and $H_2$ 
are conjugate, which gives only an isometric pair.

Therefore it would become a natural question 
 to inquire about the relation between 
the Sunada's construction of isospectral domains and the 
transplantation method.  Numerical enumeration attempted in section 5 
elucidated that at least up to $N = 13$ all the transplantable pairs 
are those pairs derived by Sunada's method. This suggests 
the possibility of its equivalence.  

The condition $n^P(\gamma) = n^Q(\gamma)$ for all possible 
sequences of $\gamma$, which is equivalent to transplantability 
of eigenfunctions, provides not only an efficient algorithm 
to enumerate transplantable pairs of planar domains, 
but has an interesting $\lq\lq$geometrical" interpretation: 
Each $n^P(\gamma)$ plays a role of the coefficient of Weyl expansion 
as mentioned in section 5. 

Transplantation method, on which we have relied in this paper,
is available only to the class of unfolded domains.
Whether isospectrality is equivalent to iso-length spectrality 
in more general domains is still an open problem.


\ack
One of authors (Y. O.) would like to thank Prof. H. Urakawa 
and Dr. M.M. Sano 
for their encouragements and helpful discussions.

\end{document}